\def\ang{\AA}

\def\gapprox{\lower.4ex\hbox{$\;\buildrel >\over{\scriptstyle\sim}\;$}}
\def\lapprox{\lower.4ex\hbox{$\;\buildrel <\over{\scriptstyle\sim}\;$}}

\def\ref#1   {\par\noindent\hangindent1cm {#1}}

\def\refer#1   {\par\noindent\hangindent1cm {#1}}

 \documentstyle[aaspp4]{article}        
 \input epsf
\slugcomment{}
\lefthead{ASCHWANDEN ET AL.}
\righthead{VERTICAL TEMPERATURE OF SOLAR CORONA}

\begin{document}
{\sl The Astrophysical Journal Letters (in press)}             \hfill{Accepted: 2000 April 6}

\title{         The Effect of Hydrostatic Weighting on the \\ 
		Vertical Temperature Structure of the Solar Corona 	}

\author{        Markus J. Aschwanden and Nariaki Nitta}
\affil{         Lockheed Martin Advanced Technology Center,
                Solar \& Astrophysics Laboratory,
                Dept. L9-41, Bldg.252,
                3251 Hanover St.,
                Palo Alto, CA 94304, USA;
                e-mail: aschwanden@lmsal.com}

\begin{abstract}
We investigate the effect of hydrostatic scale heights ${\lambda}(T)$ 
in coronal loops on the determination of the vertical temperature 
structure $T(h)$ of the solar corona. Every method that determines
an average temperature at a particular line-of-sight from optically
thin emission (e.g. in EUV or soft X-ray wavelengths) of a 
mutli-temperature plasma, is subject to the emission measure-weighted
contributions $dEM(T)/dT$ from different temperatures. Because most
of the coronal structures (along open or closed field lines) are 
close to hydrostatic equilibrium, the hydrostatic temperature scale
height introduces a height-dependent weighting function that causes
a systematic bias in the determination of the temperature structure
$T(h)$ as function of altitude $h$. The net effect is that the averaged
temperature seems to increase with altitude, $dT(h)/dh > 0$, even if
every coronal loop (of a multi-temperature ensemble) is isothermal  
in itself. We simulate this effect with
differential emission measure distributions observed by {\sl SERTS} for
an instrument with a broadband temperature filter such as {\sl Yohkoh/SXT}
and find that the apparent temperature increase due to hydrostatic weighting
is of order $\Delta T \approx T_0 \times h/r_{\sun}$. We suggest that this 
effect largely explains the systematic temperature increase in the upper corona 
reported in recent studies (e.g. by Sturrock et al., Wheatland et al., 
or Priest et al.), rather than being an intrinsic signature of a 
coronal heating mechanism.  
\end{abstract}


\section{ 		INTRODUCTION 				       }

Attempts to solve the elusive {\sl coronal heating problem} have been undertaken
by determining the heating function $E_H(h)$ as function of height $h$, 
inferred from the vertical temperature structure $T_e(h)$ of the solar corona.
In this context, a systematic temperature increase $T(h)$ with height $h$ 
has been reported from numerous observations of the quiet diffuse corona, 
coronal arcades, or coronal loops (Mariska \& Withbroe 1978; Kohl et al. 1980; 
Falconer 1994; Foley et al. 1996; Sturrock, Wheatland, and Acton 1996a; 1996b; 
Wheatland, Sturrock, \& Acton 1997; Fludra et al. 1999;
Priest et al. 1999; 2000). A common method that is chosen to infer the vertical
temperature structure $T_e(h)$ is the extraction of soft X-ray fluxes in different
wavelengths as function of height, say $F_1(h)$ and $F_2(h)$ from two different
wavelengths 1 and 2, and then to use the filter-ratio method $Q(h)=F_2(h)/F_1(h)$
to determine the temperature as function of height, $T(h)$, by inverting the
filter-ratio function $Q(T)$. The filter-ratio method has some obvious limitations,
such as the limited range where the function $Q(T)$ is unique and thus permits
only an inversion within this range, but the method has also some more subtle drawbacks
in the case of a multi-temperature plasma, as it exists in the
solar corona. In principle, the filter-ratio method is only exact for an
isothermal plasma, within the uniqueness range of $Q(T)$. The solar corona
consists of myriads of open and closed field lines filled with plasmas of almost
every temperature in the range of $10^4 \lapprox T \lapprox 10^7$ K,
which is usually quantified with a differential emission measure distribution $dEM(T)/dT$. 
This multi-thermal nature can cause systematic errors in the determination of an  
average vertical temperature profile $T_e(h)$, due to a systematic weighting bias of
the temperature-dependent pressure and density scale heights (Fig.1).
The purpose of this Letter is to demonstrate this systematic error in the
determination of the vertical temperature profile $T_e(h)$, for some typical 
observations of active regions and the quiet corona, using a broadband-filter instrument,
such as the {\sl Yohkoh Soft X-Ray Telescope (SXT)}.

\section{   		MODEL 				}

The soft X-ray flux measured along a given line-of-sight represents an integral
over emission measure contributions from plasmas with different temperatures,
which can be expressed by the differential emission measure distribution
$dEM(T)/dT$, where the emission measure contribution at a given temperature
$[T,T+dT]$ itself represents an integration along the line-of-sight $z$,
\begin{equation}
	\big( {dEM(T) \over dT} \big) dT = \int n_e^2(T,z) dz  \ .
\end{equation}
The flux measured by a detector $i$ is then given by the product of
the differential emission measure function $dEM(T)/dT$ with the instrumental
temperature response function $R_i(T)$, 
\begin{equation}
	F_i = \int {dEM(T) \over dT} R_i(T) dT \ .
\end{equation}
We characterize now the solar corona by a superposition of many different
flux tubes (along open or closed magnetic field lines), 
each one having its own temperature and density function. 
For the purpose of this demonstration we make the simplest assumption
that is compatible with observations, namely (1) that each flux tube is
near-isothermal (as it has been established for many observed EUV loops
in the temperature range of $T_e \approx 1.0-2.0$ MK,
e.g. Neupert et al. 1998; Lenz et al. 1999; Aschwanden et al. 1999;
2000a; 2000b), and that each flux tube is in near-hydrostatic equilibrium
(a condition that has been verified for EUV loops within factors of 
$\approx$1-3, Schrijver et al. 1999; Aschwanden et al. 1999; 2000a; 2000b). 
Thus, the density structure of a (near-isothermal) fluxtube can be approximated by
\begin{equation}
	n_e(h,T_e) = n_{e0} \ \exp[ - {h \over \lambda(T_e) }]   \ . 
\end{equation}
where the density (or pressure) scale height $\lambda (T_e)$ 
in hydrostatic equilibrium is proportional to the temperature $T_e$,
\begin{equation}
    {\lambda}(T_e) = { k_B T_e \over \mu m_p g_{\sun} } = {\lambda}_0 ({T_e \over 1\ {\rm MK}})  
\end{equation}
with ${\lambda}_0=47$ Mm for coronal conditions, with $\mu m_p$ the average ion mass 
(i.e. $\mu \approx 1.4$ for H:He=10:1), and $g_{\sun}$ the solar gravitation.
The differential emission measure $dEM(T,h)/dT$ is proportional to $n_e(h)^2$, and thus 
has approximatly an exponential height dependence with a scale height half of the density
scale height. In first order, the height dependence of the line-of-sight integrated
emission measure of plasma with a particular temperature $T$ can then be characterized
by the half density scale height, neglecting the curvature of the corona.
The fluxes $F_1(h)$ and $F_2(h)$ recorded in 
two detectors (i=1,2) with different wavelengths, characterized by the temperature 
response functions $R_1(T)$ and $R_2(T)$, is then
\begin{equation}
	F_i(h) = \int {dEM(T,h=0) \over dT} \ \exp[-{2 h \over {\lambda}_0 T}] \ R_i(T) dT \ ,
\end{equation}
When the filter-ratio method is applied, one takes the flux ratio of the two fluxes 
at every pixel (along a chosen altitude path $h$)
\begin{equation}
	Q(h) = {F_1(h) \over F_2(h)} \ ,
\end{equation}
which is now height-dependent, so that the resulting filter-ratio temperature
$T(Q) = T(Q[h])$ yields the height dependence of the temperature, $T(h)$.

\section{	OBSERVATIONS AND SIMULATION				}

Some typical differential emission measure distributions $dEM(T)/dT$ have been
determined with the NASA/GSFC {\sl Solar EUV Rocket Telescope and Spectrograph
(SERTS)}, using densitiy-sensitive line ratios from 8 different ionization states 
of iron between Fe$^{+9}$ (Fe X) and Fe$^{+16}$ (Fe XVII), during two flights in
1991 and 1993 (Brosius et al. 1996). These line ratios provide density diagnostics
between temperatures of $log(T_e)=5.0$ and $log(T_e)=6.7$ (i.e. $T_e \approx 0.1-5.0$ MK).
Brosius et al. (1996, see their Fig.8 and 9) derived a
differential emission measure curve $dEM(T)/dT$ in the temperature
range of $log(T_e)=4.8-7.0$, which is reproduced in Fig.2 (top panel), for two
observations of active regions (AR93, AR91) and two observations of Quiet Sun
regions (QR93, QR91).

We consider now the instrumental response functions of {\sl Yohkoh SXT}. For active
regions, the two filters sensitive to the lowest temperatures are the thin alumninium
(Al 1265 \ang ) and the Al/Mg/Mn composite filter (Tsuneta et al. 1991). 
The corresponding response functions $R_1(T)$ and $R_2(T)$ are shown in Fig.2 
(second panel), and their filter ratio $Q(T)=R_2(T)/R_1(T)$ is
given in Fig.2 (bottom panel). In order to understand the temperature contributions
to the observed flux we show the differential soft X-ray flux 
$dF(T)/dT=[dEM(T)/dT]\ R(T)$ (Fig.1, third panel), for both filters and
for all 4 regions. The differential soft X-ray flux exhibits a peak at a temperature
of $T_e \approx 10^{6.65}=4.5$ MK for the active regions, and at 
$T_e \approx 10^{6.3}=2.0$ MK for the quiet Sun regions. 

We calculate now the fluxes $F_1(h)$ and $F_2(h)$ in the two filters as function 
of the height $h$ above the limb, using the hydrostatic distribution defined in Eqs.5-6,
where each fluxtube with (different) temperature $T$ has a (different) density scale height of 
${\lambda}={\lambda}_0 T$,
while the total ensemble of fluxtubes is summed up by an integration over the entire
temperature range (i.e. temperature integral in Eqs.5-6).
The resulting SXR fluxes as function of height are shown in Fig.3 top,
illustrating that the SXR flux drops exponentially with height. We derive now the
filter ratio $Q(h)=F_2(h)/F_1(h)$, shown in Fig.3 (middle panel) for all 4 regions.
The filter ratio $Q(h)$ clearly varies as function of height $h$, although each fluxtube
is assumed to be isothermal. 

We demonstrate now what effect this filter ratio variation
$Q(h)$ has on the inference of a single-temperature model $T(h)$, as it is assumed
in the classical filter-ratio method by definition.
To invert the filter ratio $Q(T)$ as function of the temperature $T$, 
we find the following analyical approximation (accurate within $\lapprox 0.7\%$) in the
temperature range of $T=1.5-6.0$ MK (see fit in Fig.2 bottom), 
\begin{equation}
	Q(T) := {R_2(T) \over R_1(T)} \approx 0.39 + 0.27 [log(T)-6.18]^{1/2} \ .
\end{equation}  
This analytical approximation allows us conveniently to invert the filter-ratio 
temperature in the range of $Q=0.4-0.6$, i.e.
\begin{equation}
	log(T[Q]) = 6.18 + \big({Q-0.39 \over 0.27}\big)^2 \ .
\end{equation} 
The inverted temperatures $T[Q(h)]$ are shown in Fig.3 bottom for all 4 regions.
The filter ratio temperature $T(h)$ shows a height dependence from 
$T(h=0)\approx 2.1$ MK to $T(h=0.5 r_{\sun})\approx 3.1$ MK for the quiet regions, and
from 
$T(h=0)\approx 4.1-4.4$ MK to $T(h=0.5 r_{\sun})\approx 5.4-6.3$ MK for the active
regions. Thus, the weighting effect of temperature scale heights over the broadband
response function introduces an apparent temperature gradient of $dT/dh \approx 0.003$
K m$^{-1}$ for the quiet corona regions, and about $dT/dh \approx 0.005$ K m$^{-1}$
for active regions.
This corresponds about to a doubling of the apparent temperature over a distance of a
solar radius $r_{\sun}$,
\begin{equation}
	\Delta T^{SXT} \approx T_0 ({h \over r_{\sun}})	\ .
\end{equation}

\section{       	DISCUSSION AND CONCLUSIONS  		               }

We have investigated the effect of hydrostatic density scale heights in coronal
loops on the inference of a filter-ratio temperature from a broadband instrument,
in particular for the two thinnest filters of {\sl Yohkoh/SXT}, which are generally
used to derive electron temperatures in active regions and in the quiet corona.
The principal effect is that, with increasing altitude $h$ (above the solar surface),
the emission measure-weighted temperature $T_e$ becomes systematically more weighted
by the larger scale heights ${\lambda}$ (Fig.1), which are associated with loops of higher
temperature, and thus mimic an average temperature increase with height. We used
differential emission measure distributions $dEM(T)/dT$ that have been observed 
in active regions and in quiet Sun regions and simulated the temperature bias on $T(h)$ 
for the instrumental response functions of {\sl Yohkoh/SXT}. The resulting
temperature bias can be quantified approximately as 
$\Delta T^{SXT} \approx T_0 ({h / r_{\sun}})$.
We discuss now the consequences of this result.

The radial variation of temperature in the inner corona (out to 0.7 and 0.95 solar
radii) has been examined for the diffuse corona from long-exposure {\sl Yohkoh/SXT}
images by Wheatland, Sturrock, \& Acton (1997). These authors find a systematic
temperature increase from $T_e \approx 1.6$ MK near the solar surface to $T_e \approx 2.4$
at a height of 0.5 solar radii for the 7-9 May 1992 active region, and from
$T_e\approx 1.8$ MK to $T_e\approx 2-3$ MK at 1 solar radius for the 26 August 1992 region.  
This systematic temperature increase of the solar corona was interpreted in terms
of a downward heat flux, leading to the conclusion of a heat deposition above the
observed height. According to our model (Eq.9), we estimate fully consistent temperature
increases [$T_e(h=0)=1.6$ MK $\mapsto T_e(h=0.5 r_{\sun})=2.4$ MK for the first case,
and ($T_e(h=0)=1.8$ MK $\mapsto T_e(h=r_{\sun})=2.7$ MK for the second case] from
the emission measure-weighted hydrostatic scale heights alone, even if all fluxtubes
are isothermal. Therefore, if the hydrostatic weighting effect on the Yohkoh/SXT
filter ratio method would be corrected, no net temperature increase would result,
and thus no support for a heating function in the upper corona is warranted. 

With the same measurement technique, Priest et al. (1999; 2000) analyzed large-scale
arcades and loops and found a temperature increase from $T_e(h=0)=1.6$ MK to 
$T_e(h=0.5 r_{\sun})=2.2-2.3$ MK for a first loop observed on 1992 Oct 3, 
an increase from $T_e(h=0)=1.6$ MK to $T_e(h=500$ Mm$)=2.4-2.6$ MK in a second loop,
and an increase from $T_e(h=0)=1.6$ MK to $T_e(h=350$ Mm$) \approx 2.1$ MK in a third loop.
The authors fitted three heating models to these temperature increases $T_e(h)$ and
found that a uniform heating function provides the best fit for all 3 cases, while
heating functions localized at the loop top was found to be less likely, and a
heating function localized near the loop footpoints was rejected. From our
model (Eq.9) we can reproduce the same temperature increases for these 3 cases,
so that virtually no net temperature increase remains, if
the {\sl Yohkoh/SXT} filter ratios would be corrected for the hydrostatic
emission measure weighting. Comparing the corrected temperature profiles
$T_e(h) \approx const$ with the heating models shown of Figs.8 and 9
in Priest et al. (2000), one would conclude that the data are most consistent
with the theoretical model of footpoint heating, 
a conclusion that would also be more in line with other
recent observations from {\sl TRACE} (Schrijver et al. 1999; Aschwanden et al. 2000b).

In summary, we like to point out that filter ratio temperatures from broadband
instruments may lead to systematic errors in the determination of vertical
temperature profiles $T_e(h)$, that can only be corrected properly by forward-fitting
of models which contain both temperature $T_e(h)$ and density profiles $n_e(h)$.
The systematic effects are larger for broadband filter ratios (e.g. {Yohkoh/SXT}) 
than for narrowband filters (e.g. {\sl SoHO/EIT} or {\sl TRACE}). Any detected
temperature increase derived from an emission measure-weighted temperature definition 
is subject to the {\sl hydrostatic weighting of a multi-temperature plasma}, and does 
not directly describe a variation ($dT/dh$) of the electron temperature along a 
magnetic field line.

\bigskip
{\sl Acknowledgements:} 
We appreciate helpful comments from Drs. David Alexander, Jim Lemen,
Karel Schrijver, Greg Slater, and the referee. This work was supported by 
{\sl Yohkoh/SXT} NASA contract NAS8-40801. 

\clearpage


\section*{              References                                              }

\refer{Aschwanden,M.J., Newmark,J.S., Delaboudiniere,J.P., Neupert,W.M.,
        Klimchuk,J.A., G.A.Gary, Portier-Fozzani,F., and Zucker,A. 1999, ApJ 515, 842}
\refer{Aschwanden,M.J., Hurlburt,N., Alexander,D., Newmark,J.S., Neupert,W.M.,
    	Klimchuk,J.A., and G.A.Gary 2000a, ApJ 531, (March 1 issue), in press}
\refer{Aschwanden,M.J., Nightingale,R., Alexander,D., and Reale,F. 2000b, ApJ, subm.}
\refer{Brosius,J.W., Davila,J.M., Thomas,R.J., and Monsignori-Fossi,B.C. 1996,
	ApJS 106, 143}
\refer{Falconer,D. 1994, Relative Elemental Abundance and Heating Constraints
	Determined for the Solar Corona from SERTS Measurements, NASA Tech.Memo.
	104616}
\refer{Feldman,U., Doschek,G.A., Sch\"uhle,U., and Wilhelm,K. 1999,
	ApJ 518, 500}
\refer{Fludra,A., DelZanna,G., Alexander,D., and Bromage,B,J.I. 1999,
	JGR 104/ No.A5, 9709}
\refer{Foley,C.A., Acton,L.W., Culhane,J.L., \& Lemen,J.R. 1996, IAU Colloq. 153,
	Magnetohydrodynamic Phenomena in the Solar Atmosphere, p.419}
\refer{Kohl,J., Weiser,H., Withbroe,G., Noyes,R., Parkinson,W., Reeves,E.,
	Munro,R., and MacQueen,R. 1980, ApJ 241, L117}
\refer{Lenz,D., DeLuca,E.E., Golub,L., Rosner,R., and Bookbinder,J.A. 1999, ApJ 517, L155}
\refer{Mariska,J. and WIthbroe,G. 1978, Sol.Phys. 60, 67}
\refer{Neupert,W.M. et al. 1998, SP 183, 305}
\refer{Priest,E.R., Foley,C.R., Heyvaerts,J., Arber,T.D., Culhane,J.L.,
	and Acton,L.W. 1999, Nature, Vol.393, No.6685 (June 11 issue).} 
\refer{Priest,E.R., Foley,C.R., Heyvaerts,J., Arber,T.D., Mackay,D., Culhane,J.L.,
	and Acton,L.W. 2000, subm.}
\refer{Schrijver,C.J. et al. 1999,  Solar Phys. 187, 261}
\refer{Sturrock,P.A., Wheatland,M.S., and Acton,L.W. 1996b, ApJ 461, L115} 
\refer{Sturrock,P.A., Wheatland,M.S., and Acton,L.W. 1996a, IAU Colloq. 153,
	Magnetohydrodynamic Phenomena in the Solar Atmosphere, p.417}
\refer{Tsuneta,S. et al. 1991, Sol.Phys. 136, 37}
\refer{Wheatland,M.S., Sturrock,P.A., and Acton,L.W. 1997, ApJ 482, 510}

\clearpage


\section*{              Figure Captions                                  }

{\bf Fig.1: }{This cartoon illustrates scale height-weighted contributions of
hydrostatic loops or open fluxtubes to the emission measure observed along two
line-of-sights above the solar limb. The left line-of-sight at a height of
$h=100$ Mm above the limb samples significant emission from the 3 loops with
temperatures of 1.5-2.5 MK. The right line-of-sight at a height of $h=200$ Mm
above the limb samples significant emission only from the hottest loop with
$T=2.5$ MK.}

{\bf Fig.2: }{The differential emission measure distribution $dEM(T)/dT$ of two
active regions (AR93, AR91) and two quiet Sun regions (QR93, QR91) measured by
Brosisus et al. (1996) with SERTS data (top panel). The Yohkoh/SXT response
function for the two thinnest filters (second panel). The differential SXR fluxes
$dF(T)/dT=[dEM(T)/dT]*R(T)$ for the two SXT filters (thin and thick linestyles)
for all 4 regions (third panel). The filter ratio $Q(T)=R_2(T)/R_1(T)$ for the
two Yohkoh/SXT filters and an analytical approximation in the range of
$T=1.5-6.0$ MK (bottom panel).}

{\bf Fig.3:}{The height dependence of the observed SXT fluxes $F(h)$ for the
two filters (thin and thick linestyles) and all 4 regions (different linestyles)
(top panel). The resulting filter ratio $Q(h)$ for all 4 regions (second panel),
and the inferred filter-ratio temperatures (bottom panel). Note that the
filter-ratio temperature $T(h)$ shows a systematic increase with height,
although a model with isothermal loops was assumed.}

\clearpage
\plotone{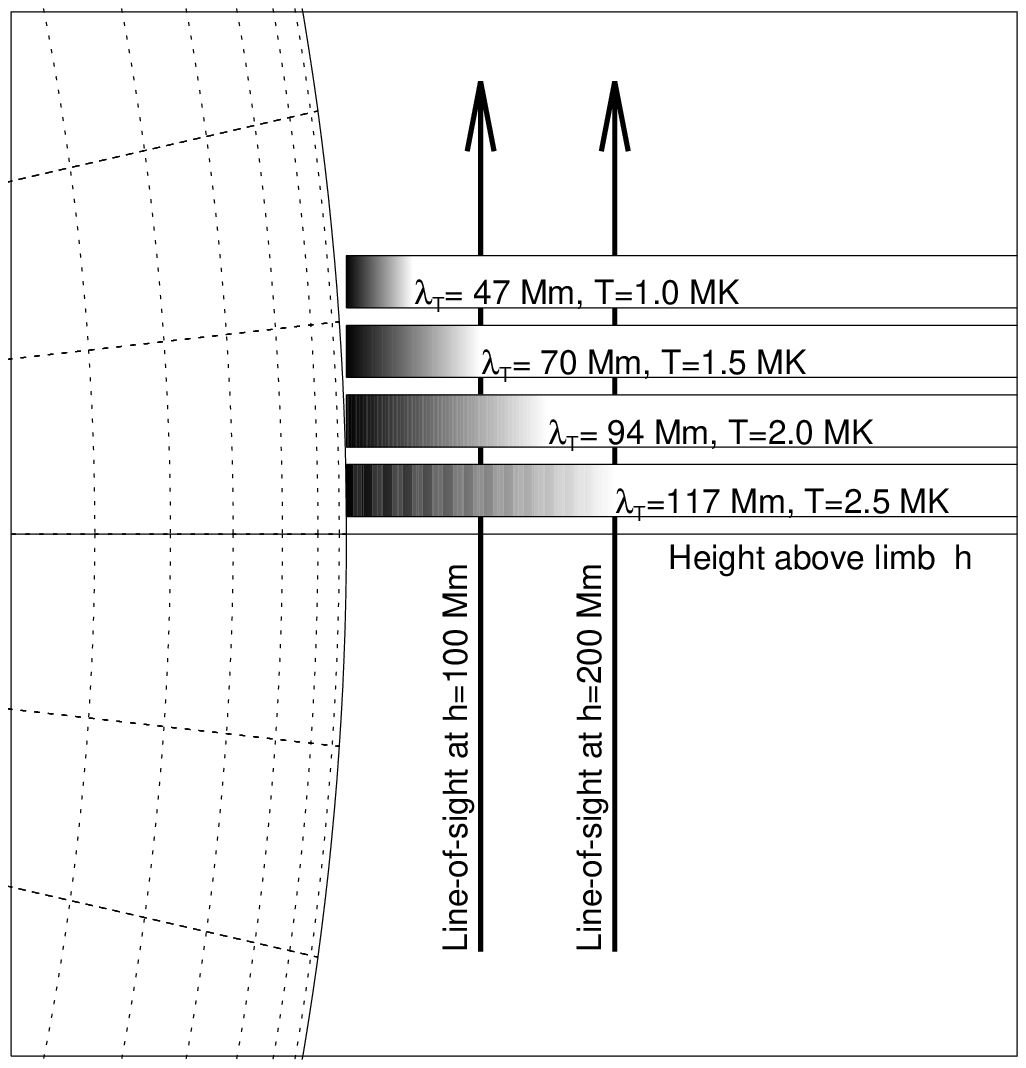}
\clearpage
\plotone{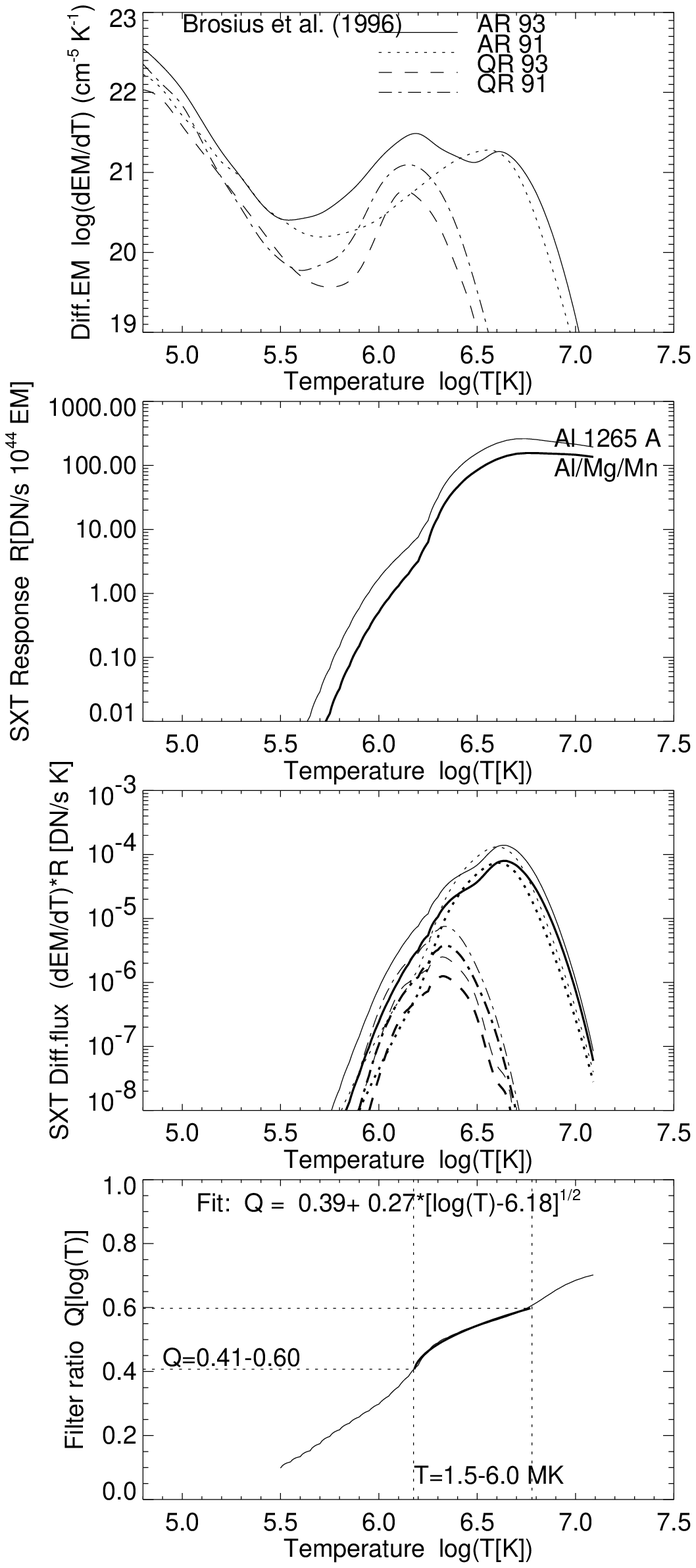}
\clearpage
\plotone{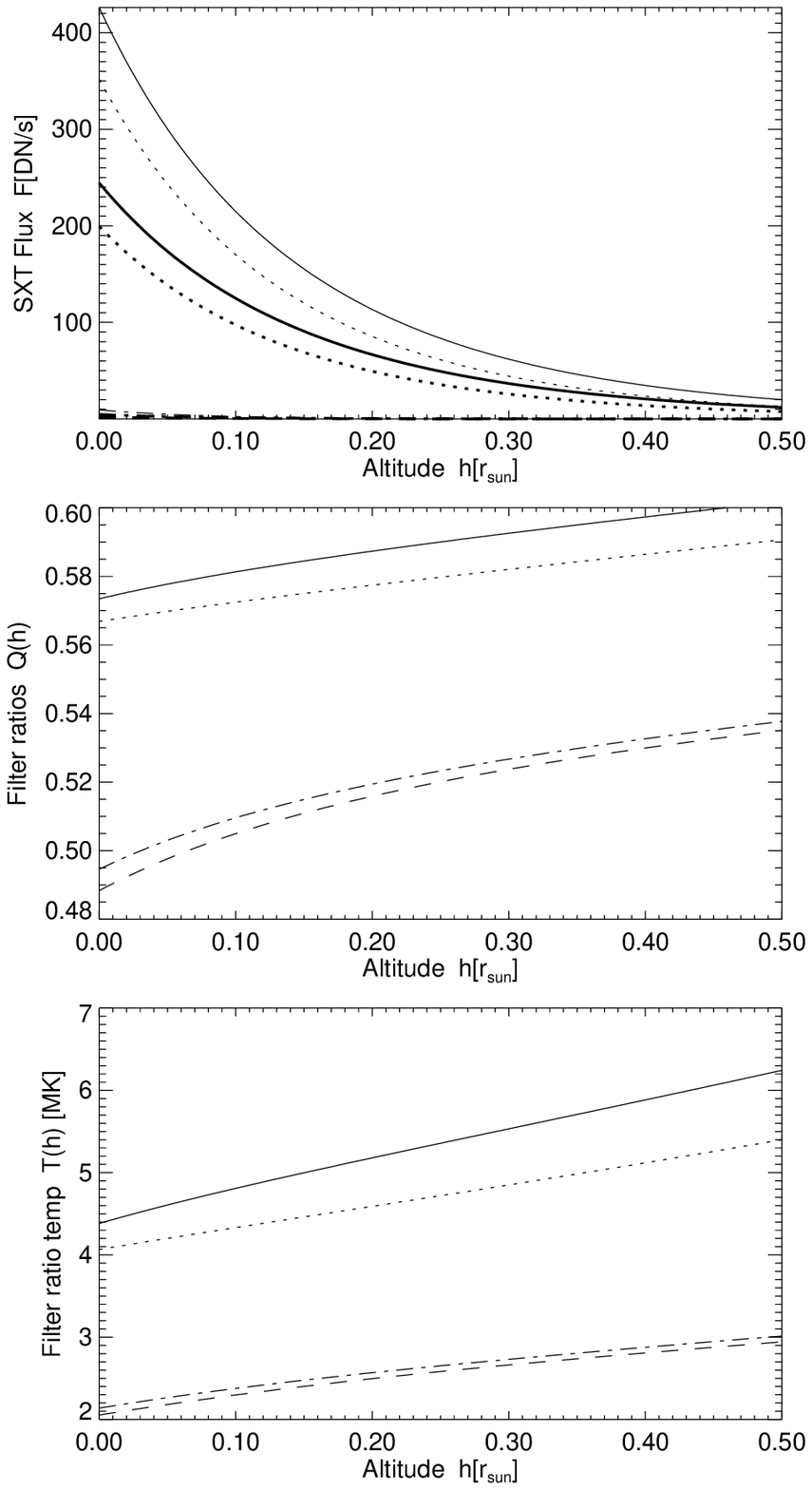}

\end{document}